\documentclass[12pt,preprint]{aastex}
\usepackage{graphicx}
\usepackage{natbib}

\shorttitle{Gravitational Lensing of stars orbiting Sgr A*}
\shortauthors{Bozza and Mancini}

\begin{document}

\title{Gravitational Lensing of stars in the central arcsecond of our Galaxy}

\author{V. Bozza$^{1,2,3}$ and L. Mancini$^{2,3}$}
\affil{$^1$Centro Studi e Ricerche ``Enrico Fermi'', via
Panisperna 89/A, Roma, Italy.}

\affil{$^2$Dipartimento di Fisica ``E.R. Caianiello'',
Universit\'a di Salerno, via S. Allende, Baronissi (SA), Italy.}

\affil{$^3$Istituto Nazionale di Fisica Nucleare, Sezione di
Napoli, Italy.}

\begin{abstract}
In the neighborhood of Sgr A*, several stars (S2, S12, S14, S1,
S8, S13) enjoy an accurate determination of their orbital
parameters. General Relativity predicts that the central black
hole acts as a gravitational lens on these stars, generating a
secondary image and two infinite series of relativistic images.
For each of these six stars, we calculate the light curves for the
secondary and the first two relativistic images, in the
Schwarzschild black hole hypothesis, throughout their orbital
periods. The curves are peaked around the periapse epoch, but two
subpeaks may arise in nearly edge-on orbits, when the source is
behind or in front of Sgr A*. We show that for most of these stars
the secondary image should be observable during its brightness
peak. In particular, S14 is the best candidate, since its
secondary image reaches K=23.3 with an angular separation of 0.125
mas from the apparent horizon of the central black hole. The
detection of such images by future instruments could represent the
first observation of gravitational lensing beyond the weak field
approximation.
\end{abstract}


\section{Introduction}

In the last years, the existence of a supermassive black hole
corresponding to the radio source Sgr A* has been established on a
quite solid ground \citep{MelFal}. The studies of stellar dynamics
in the neighborhood of Sgr A* have played a major role in this
scientific process providing strong evidence in favor of the
existence of a huge point-like mass hidden in this radio source.
The most accurate observations have been led by an American group
at Keck \citep{Ghez1,Ghez2,Ghez3}, and by a German group at VLT
\citep{EckGen,Genzel,Eckart,Sch1,Sch2,Eis}.

The measurement of the proper motions of stars very close to the
central black hole has led to a three-dimensional reconstruction
of the orbits of six stars, named S2, S12, S14, S1, S8, S13 by
\citet{Sch2} and  S0-2, S0-19, S0-16, S0-1, S0-4, S0-20  by
\citet{Ghez3}, respectively. The ambiguity on the sign of the
inclination was resolved for the first of these stars by Doppler
shift measures \citep{Ghez2}. Then, very recently, further
measurements have resolved this ambiguity also for the remaining
stars \citep{Eis}. Now the orbital parameters of all of them are
known to an accuracy reasonable enough to start investigations
relying on the knowledge of the geometric configurations between
the star and the central black hole. Indeed one of the most
interesting effects to search for is gravitational lensing.

The bending of light by gravitational fields is an invaluable
instrument to investigate the properties of the lensing object,
its mass distribution, the gravitational potential and many
general relativistic effects. The stars around the black hole at
the Galactic Center provide a unique lab for gravitational lensing
researches, which may have very much to say in the near future on
gravitational physics.

The first studies of gravitational lensing by Sgr A* have been
performed by \citet{WarYus}, who focused on sources in the stellar
cluster surrounding it. They realized that gravitational lensing
would slightly deplete the sky of visible stars in a region of
radius 10 mas, pushing them to 30 mas. However, such a statistical
effect is practically unobservable given the observed stellar
density in the neighborhood of Sgr A*. Gravitational lensing could
be recognized by identifying opposite images of a star passing
behind the central black hole. Several stars should be
significantly lensed at a given time. Estimates by \citet{Jar},
\citet{AleSte}, \citet{CGM} agree that roughly 10 stars in the
Galactic bulge are simultaneously lensed by Sgr A* with a
detection threshold for the secondary image of 23 mag in the
K-band. However, recognizing gravitational lensed images is by no
way a trivial task. While the secondary image would stay very
close to the central black hole (at about 30 mas), the direct
image of the source could be displaced by as much as several
arcsecs. The association between the two images could be
established on the basis of common spectral properties, flux
variation and full correlation in their proper motions (the
lensing event should have a duration of some months). Eventual
observations of several pairs of images could pinpoint the exact
position of the black hole \citep{Ale}. The secondary images could
be additionally lensed by a possible population of stellar size
black holes around the supermassive one \citep{AleLoe,CGM,Muno},
though such a double microlensing event is relatively unlikely.
Another possibility is to measure the shift of the apparent
position of a star passing just behind the black hole
\citep{NusBro}. Such a measurement, would provide an independent
estimate of the distance of Sgr A*.

Coming to our six stars, we have an evident advantage: we
precisely know their location in space as a function of time and
thus the geometry of the lensing configuration. Since we already
observe the direct image, we can predict at any time where to look
for a secondary lensing image and what its apparent magnitude
should be. Thanks to our knowledge of their orbits, we can easily
predict the best time to observe these secondary images. Finally,
since the distances between the sources and the lens are known,
the information extraction from gravitational lensing observations
is much easier and unambiguous.

A very important fact that must be taken into account in the study
of the light curves of gravitational lensing images is that the
weak field regime for the secondary image holds only for a very
good alignment between source, lens and observer. This is not the
case for the six stars studied in this article. This forces us to
resort to the full General Relativity deflection angle formula,
which is very well known since old times in terms of elliptic
integrals \citep{Dar}. The deflection angle diverges
logarithmically at a characteristic minimum impact angle for any
class of spherically symmetric black holes \citep{Boz1}, implying
the existence of two infinite series of "ghosts", as called by
\citet{Dar}, or relativistic images, as named by \citet{VirEll}.
These images are due to photons winding one or more times around
the black holes and are usually extremely faint. Moreover, they
would appear very close to the central black hole, at angular
separations comparable to the apparent Schwarzschild horizon,
which is about 23 $\mu$as. The detection of such images would thus
require a higher order technological effort, including very long
baseline interferometry in the K-band and very high sensitivity.
However, these images are very precious witnesses of the extreme
gravitational field at distances slightly larger than the
Schwarzschild radius of the central black hole. The information
stored in them is of striking importance to unveil the features of
the true gravitational theory and the mysteries on the physics of
the central black hole. Relativistic images produced by Sgr A*
have also been studied by \citet{Pet}, who noticed that their
contribution to microlensing is negligible, and by \citet{EirTor}
in the context of retro-lensing.

S2 was the first star around Sgr A* to be studied as a possible
source for gravitational lensing effects \citep{DeP}. Even if its
orbit never allows a good alignment with Sgr A*, the secondary
image can reach K$\lesssim 30$ while it is in a fully relativistic
regime. In a previous work \citep{BozMan}, we presented a complete
study of S2 gravitational lensing using the tools of the Strong
Field Limit method \citep{BCIS,Boz1}. In this way we have been
able to predict the precise epoch of the best observability of the
secondary image.

In this paper we develop a complete analysis of the six
aforementioned stars as potential sources for gravitational
lensing by the central black hole, finding interesting results,
especially for the star S14. To cover the whole range of
gravitational lensing regimes, from strong to weak, we will
exclusively use the exact deflection formula known for the
Schwarzschild black hole. In this way we can achieve a very
reliable and precise description for the light curves of all
images.

Our paper is structured as follows. In \S~ 2 we review the exact
Schwarzschild deflection formula and some basics of gravitational
lensing. In \S~ 3 we recall the features of the six stars we want
to examine, discussing the geometry of their orbits. In \S~ 4 we
present the results of our analysis, with the light curves of our
six stars and the positions of the secondary images. In \S~ 5 we
discuss the perspectives for observations of these images using
long baseline interferometry. Finally, \S~ 6 contains the
conclusions.

\section{Schwarzschild gravitational lensing}

\begin{figure}
\resizebox{\hsize}{!}{\includegraphics{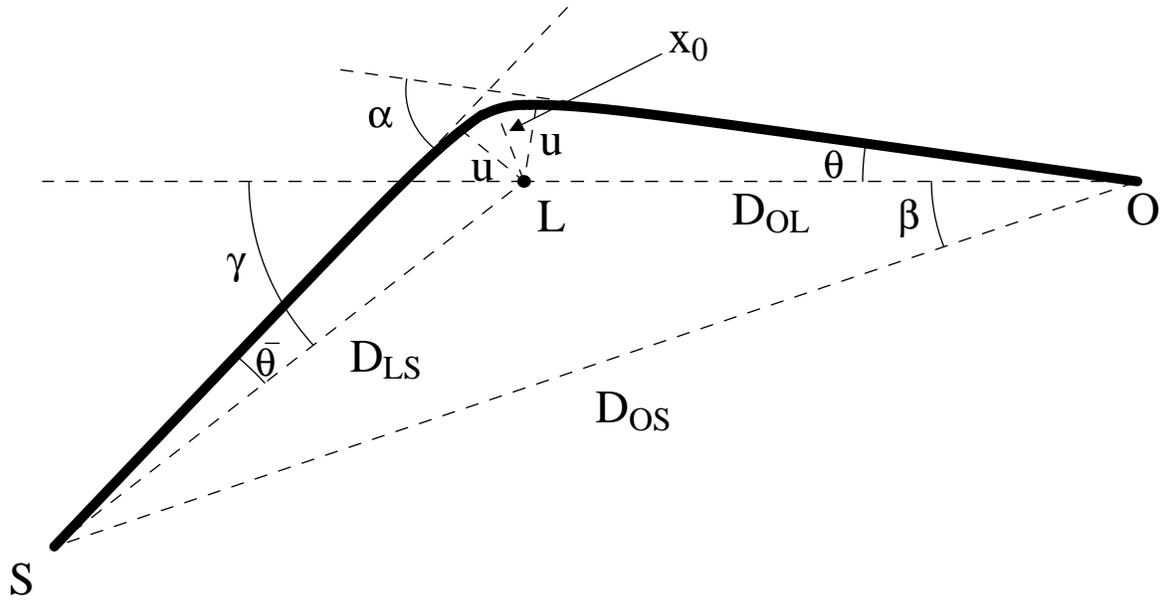}}
 \caption{A generic lensing configuration.}
 \label{Fig Lens Eq}
\end{figure}

Consider the generic three-point configuration between source,
lens and observer, illustrated in Figure \ref{Fig Lens Eq}. We
indicate by $D_{OL}$, $D_{LS}$ and $D_{OS}$ the distance between
observer and lens, lens and source, observer and source
respectively. All the distances are measured in terms of the
Schwarzschild radius of the black hole
\begin{equation}
R_{Sch}=\frac{2G M}{c^2}.
\end{equation}

We call optical axis the line connecting the observer and the
lens. We denote by $\gamma$ the angle between the line connecting
the source to the lens and the optical axis, and by $\beta$ the
angle formed by the line connecting source and observer with the
optical axis. The direct image of the source suffers (weak)
gravitational lensing only when $\gamma$ is very close to zero.
The secondary image is formed by photons passing behind the black
hole, as shown in Figure \ref{Fig Lens Eq}. The impact angles for
the photons as seen from the observer and the source are $\theta$
and $\overline{\theta}$, respectively. The deflection angle is
defined as the angle between the asymptotic directions of motion
of the photons, before and after the encounter with the black
hole.

By inspection of Figure \ref{Fig Lens Eq}, we can write down the
following lens equation
\begin{equation}
\gamma=\alpha(\theta)- \theta -\overline{\theta}. \label{Lens Eq}
\end{equation}

The impact parameter $u$ is related to the impact angles $\theta$
and $\overline{\theta}$ by
\begin{equation}
u= D_{OL} \sin\theta =D_{LS} \sin \overline{\theta}.
\end{equation}
This relation can be used to eliminate $\overline{\theta}$ in
favor of $\theta$. The position angle of the source $\gamma$ runs
from 0 (perfect alignment) to $\pi$ (perfect anti-alignment).

Besides the direct and the secondary image, we also have higher
order relativistic images, which form closer to the central black
hole. Because of their potential importance as probes of very
strong gravitational fields, in our analysis we will also discuss
the third and the fourth order image. The third order image is
formed by photons turning around the black hole and reaching the
observer from the same side of the direct image. It is described
by a lens equation of the type
\begin{equation}
2\pi-\gamma=\alpha(\theta)- \theta -\overline{\theta}.
\end{equation}
Notice that the deflection angle $\alpha(\theta)$ is in the range
$[\pi, 2\pi ]$.

The fourth order image is formed by photons performing a complete
loop around the black hole and reaching the observer from the side
of the secondary image. It is described by a lens equation of the
type
\begin{equation}
2\pi+\gamma=\alpha(\theta)- \theta -\overline{\theta}.
\end{equation}
The required deflection angle is in the range $[2\pi, 3\pi ]$.

The direct and the third order image have positive parity while
the secondary and the fourth order ones have negative parity.

The exact deflection angle was calculated by \citet{Dar} in terms
of elliptic integrals. It can be written as a function of the
closest approach distance $x_0$ as

\begin{equation}
\alpha(x_0)=-\pi-4 F \left( \varphi_0,\lambda
\right)G\left(x_0\right), \label{alpha exact}
\end{equation}
where
\begin{equation}
G\left(x_0\right)=\sqrt{\frac{8x_0\left(-3+x_0+\sqrt{(-1+x_0)(3+x_0)}
\right)}{3-2x_0}}
\end{equation}
and
\begin{equation}
F \left( \varphi_0,\lambda
\right)=\int\limits_0^{\varphi_0}\left(1-\lambda \sin^2 \varphi
\right)^{-1/2} {d} \varphi%
 \label{Elliptic function}
\end{equation}
is an elliptic integral of first kind. The parameters $\varphi_0$
and $\lambda$ are themselves functions of $x_0$:
\begin{equation}
\varphi_0=\arcsin \sqrt{\frac{-3+x_0-\sqrt{-3+2x_0+x_0^2}}
{2\left(-3+2x_0 \right)}} ,
\end{equation}
and
\begin{equation}
\lambda=\frac{3-x_0-\sqrt{-3+2x_0+x_0^2}}
{3-x_0+\sqrt{-3+2x_0+x_0^2}}.
\end{equation}

The relation between the closest approach distance $x_0$ and the
impact parameter $u$ can be easily found by the conservation of
angular momentum
\begin{equation}
x_0^2=\left(1-\frac{1}{x_0} \right) u^2.
\end{equation}

The use of the exact formula (\ref{alpha exact}) is rather
cumbersome in analytical applications. Actually, before Darwin's
work only the weak field limit for large impact parameters was
known. In Schwarzschild units, it reads
\begin{equation}
\alpha_{WFL}= \frac{2}{u}.
\end{equation}

The opposite limit is recovered when $u$ approaches its minimum
value $u_m=3\sqrt{3}/2$. In this limit, we have
\citep{Dar,Oha,BCIS}
\begin{equation}
\alpha_{SFL}=-\log\left( \frac{u}{u_m}-1
\right)+\log\left[216(7-4\sqrt{3}) \right]-\pi.
\end{equation}
This formula was later extended by \citet{Boz1} to a generic
spherically symmetric black hole and to a Kerr black hole for
equatorial motion \citep{Boz2}. The logarithmic divergence of the
deflection angle at the minimum impact parameter signals the
presence of an infinite sequence of images with deflection angles
which differ by a multiple of $2\pi$ (an arbitrary number of loops
around the black hole). For impact parameters below $u_m$, the
photon is always captured by the black hole. The striking
importance of higher order images comes from the fact that they
can be directly related to the values of the metric components and
their derivatives close to the horizon \citep{Boz1}. This provides
a unique way to identify the class to which the black hole
belongs, possibly giving hints on the correct gravitational theory
that holds in the strong field regime.

\begin{figure}
\resizebox{\hsize}{!}{\includegraphics{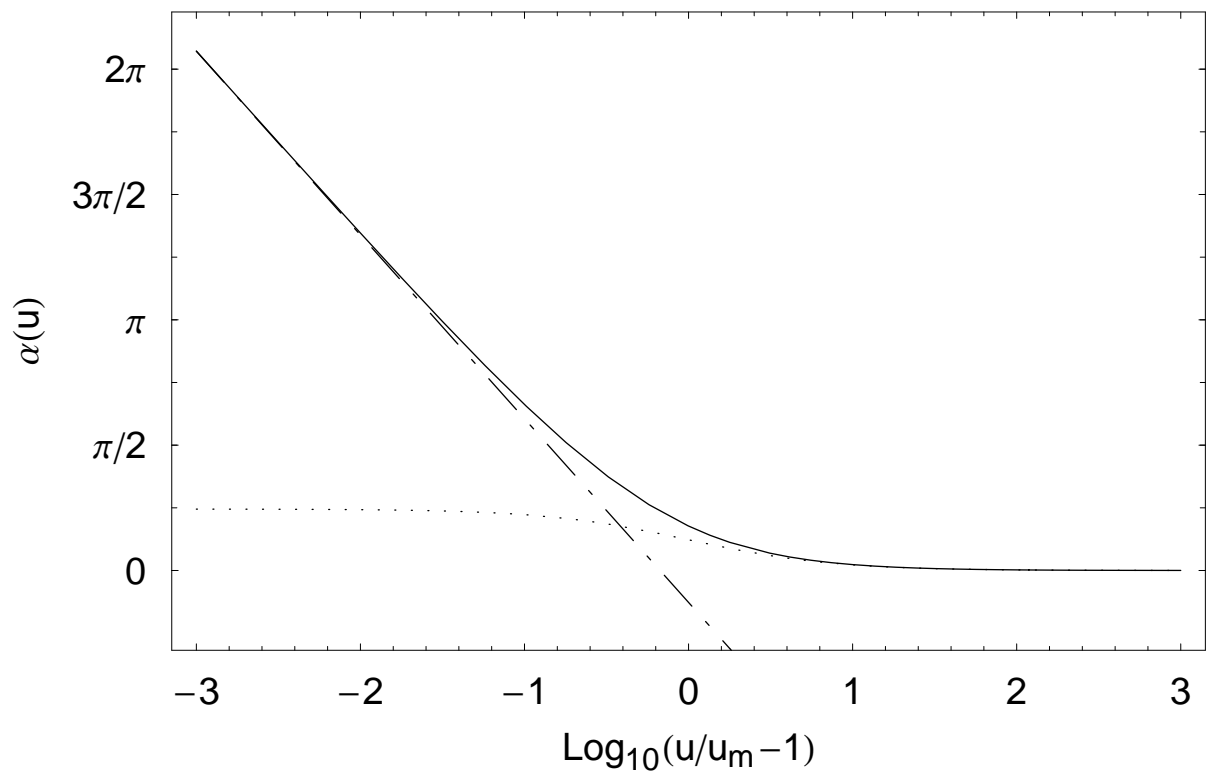}}
 \caption{Comparison between the exact deflection angle (solid line),
 the weak field approximation (dotted line) and the strong field limit approximation
 (dot-dashed line), as functions of $\mathrm{Log}_{10}(u/u_m-1)$.}
 \label{Fig Compari}
\end{figure}

In Figure \ref{Fig Compari}, we compare the exact deflection angle
with the two approximations. We see that the weak field limit is
valid only for very small deflection angles at large impact
parameters, while the strong field limit starts to be accurate
from $\alpha=\pi$ at impact parameters very close to the minimum
value. For our purpose neither of the two limits is suitable,
since the secondary image demands a deflection angle in the whole
range $[0, \pi ]$.

Our strategy is to solve the exact lens equations for the direct,
the secondary, the third order and fourth order image, using the
approximate solutions as starting points (the weak field solution
for the direct and the strong field ones for the others). We shall
keep the terms $\theta$ and $\overline{\theta}$ in the equations,
since they are relevant in the weak field regime and in the
intermediate one, though they become very small in the strong
field regime, at least for the sources considered here.

The magnification of an image at angle $\theta$ is given by the
general formula
\begin{equation}
\mu=\frac{D_{OS}^2}{D_{LS}^2}\frac{\sin\theta}{ \frac{d
\gamma}{d\theta} \sin{\gamma}}. \label{mu}
\end{equation}

The derivative can be suitably approximated by the finite
difference ratio, as we do not solve the lens equation
analytically. Armed with these tools, once we know the absolute
magnitude and the position of a star around Sgr A*, we can give
the position and the apparent magnitude of its four most relevant
images.

\section{Gravitational Lensing candidates around Sgr A*}

\begin{figure}
\resizebox{13cm}{!}{\includegraphics{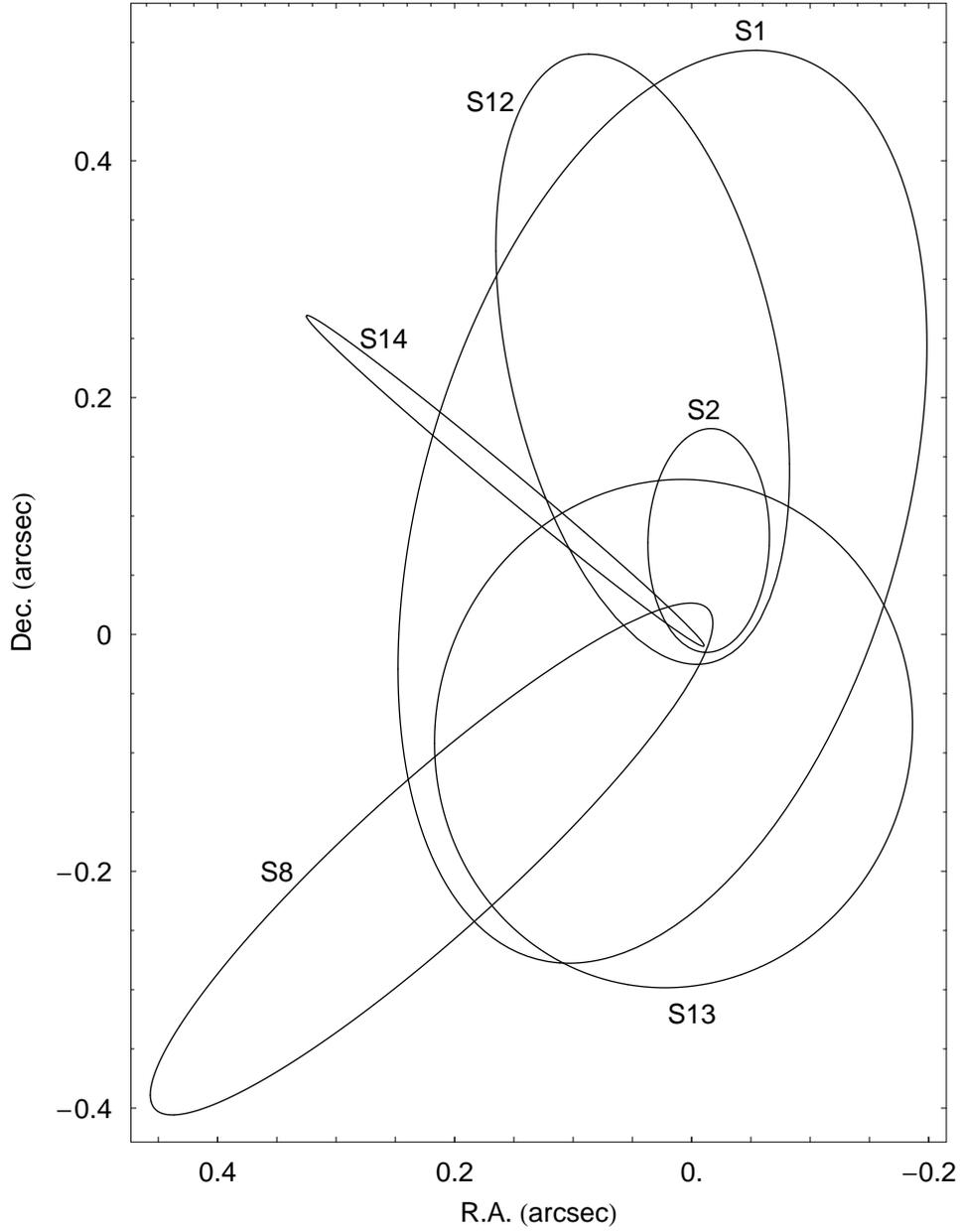}}
 \caption{Orbits of the stars S2, S12, S14, S1, S8, S13 around Sgr A*, following \citet{Eis}.}
 \label{Fig Orbite}
\end{figure}

In the neighborhood of Sgr A*, proper motion and Doppler
measurements have strongly increased our knowledge of stellar
dynamics in this region. For six stars a complete reconstruction
of the orbit has been achieved (Fig. \ref{Fig Orbite}). For our
analysis we shall follow the latest data published by \citet{Eis}.
Alternative solutions (compatible within uncertainties) for the
orbits of these six stars, derived by different measurements and
methodologies, can be found in \citet{Ghez3}.

All these stars appear to belong to early spectral types B0-B9
\citep{Eis}, opening questions about the presence of such young
stars in the inner few hundredths of a parsec around the Galactic
center. Table 1 sums up their orbital parameters, taken from
\citet{Eis}. These stars have been observed in the K-band centered
on $\lambda=2.2$ $\mu$m, in which the extinction has been
estimated to be 2.8 mag \citep{Eis}. In these new data all the
uncertainties have been drastically reduced, save for S13, whose
orbital plane is still largely undetermined. In our analysis, we
will only use the best fit values of each parameter. In the last
section we will discuss how the uncertainties may affect our
predictions. The Schwarzschild radius of the central black hole is
\begin{equation}
R_{Sch}=\frac{2GM_{BH}}{c^2}=0.071 \;\; \mathrm{ AU},
\end{equation}
taking for its mass the value $M_{BH}=3.61\times 10^6$ M$_\odot$
\citep{Eis}. The distance of the Sun from the Galactic center is
currently estimated to $D_{OL}\simeq 8$ kpc \citep{Reid}; then the
apparent size of the event horizon is
\begin{equation}
\theta_m=\frac{3\sqrt{3}R_{Sch}}{2D_{OL}}=23 \mathrm{\mu as}.
\end{equation}

Among our stars, the minimum distance from the central black hole
is reached by S14, which touches 110 AU, still 1545 times greater
than $R_{Sch}$. Hence, we can safely consider, from the
gravitational lensing point of view, both the observer and the
source as being infinitely distant from the lens. The case of a
source orbiting very close to the lens has been examined in the
context of stellar black holes by \citet{CunBar}.

\begin{table*}
\centering
\begin{scriptsize}

\begin{tabular}{ccccccc}
 \hline \hline

  Parameter & S2 & S12 & S14 & S1 & S8 & S13 \\
  \hline
  $a$ (arcsec) \dotfill & $0.1226\pm 0.0025$ & $0.286\pm0.012$ & $0.225\pm0.022$ & $0.412\pm0.024$ & $0.329\pm0.018$ & $0.219\pm0.058$ \\
  $P$ (yr) \dotfill & $15.24\pm 0.36$ & $54.4\pm3.5$ & $38.0\pm 5.7$ & $94.1\pm9.0$ & $67.2\pm5.5$ & $36\pm15$ \\
  $e$ \dotfill & $0.8760\pm 0.0072$ & $0.9020\pm0.0047$ & $0.9389\pm0.0078$ & $0.358\pm0.036$ & $0.927\pm0.019$ & $0.395\pm0.032$ \\
  $T_0$ (yr)  \dotfill& $2002.315\pm0.012$ & $1995.628\pm0.016$ & $2000.156\pm0.052$ & $2002.6\pm0.6$ & $1987.71\pm0.81$ & $2006.1\pm1.4$ \\
  $i$ (deg)  \dotfill& $131.9\pm1.3$ & $32.8\pm1.6$ & $97.3\pm2.2$ & $120.5\pm1.0$ & $60.6\pm5.3$ & $11\pm35$ \\
  $\Omega$ (deg)  \dotfill& $221.9\pm1.3$ & $233.3\pm4.6$ & $228.5\pm1.7$ & $341.5\pm0.9$ & $141.4\pm1.9$ & $100\pm198$ \\
  $\omega$ (deg)  \dotfill& $62.6\pm1.4$ & $311.8\pm3.6$ & $344.7\pm2.2$ & $129.8\pm4.7$ & $159.2\pm1.8$ & $250\pm161$ \\
  $K$ (mag) \dotfill & 13.9 & 15.5 & 15.7 & 14.7 & 14.5 & 15.8
  \\
 \hline

\end{tabular}
\end{scriptsize}

\caption{Orbital parameters of the six stars examined in the
paper: $a$ is the semimajor axis, $P$ is the orbital period, $e$
is the eccentricity, $T_0$ is the epoch of periapse, $i$ is the
inclination of the normal of the orbit with respect to the line of
sight, $\Omega$ is the position angle of the ascending node,
$\omega$ is the periapse anomaly with respect to the ascending
node (data taken from \citealt{Eis}). $K$ is the magnitude in the
K-band, taken from \citet{Sch2}.}
\end{table*}

In order to calculate the position and the brightness of the
gravitational lensing images of our stars, we need to track their
position with respect to the central black hole as a function of
time. In particular, we need their distance $D_{LS}$ and their
position angle $\gamma$ with respect to the optical axis. By
simple geometry, we can relate them to the anomaly angle from the
periapse, denoted by  $\phi$
\begin{eqnarray}
&& D_{LS}=\frac{a(1-e^2)}{1+e \cos \phi}\\ &&
\gamma=\arccos[\sin(\phi+\omega)\sin i], \label{gamma}
\end{eqnarray}
where $a$ is the semimajor axis, $e$ is the eccentricity, $i$ is
the inclination of the orbit and $\omega$ is the periapse anomaly
with respect to the ascending node. To find $\phi$ as a function
of time, we can use angular momentum conservation (neglecting
corrections by possible deviations of the mass distribution from
spherically symmetry), which provides a differential equation for
$\dot \phi$
\begin{equation}
\frac{\left[a(1-e^2) \right]^{3/2}}{\sqrt{G M_{enc}}(1+e \cos
\phi)^2}\dot \phi=1, \label{dotphi}
\end{equation}
where $M_{enc}$ is the mass enclosed in the orbit of the star we
are considering, related to the orbital period $P$ by
\begin{equation}
G M_{enc}=4\pi^2\frac{a^3}{P^2}.
\end{equation}

Integrating and inverting equation (\ref{dotphi}), we can get
$\phi$ as a function of time, exploiting the initial condition
$\phi(T_0)=0$, with $T_0$ given in Table 1.

\section{Results}

In this section we shall present the outcome of the gravitational
lensing analysis for our six stars. We will show the light curves
of the secondary, third and fourth order images, and the expected
trajectory of the secondary image with respect to the central
black hole. In the following, we shall discuss the results
individually for each of them.

\begin{figure*}
\resizebox{\hsize}{!}{\includegraphics{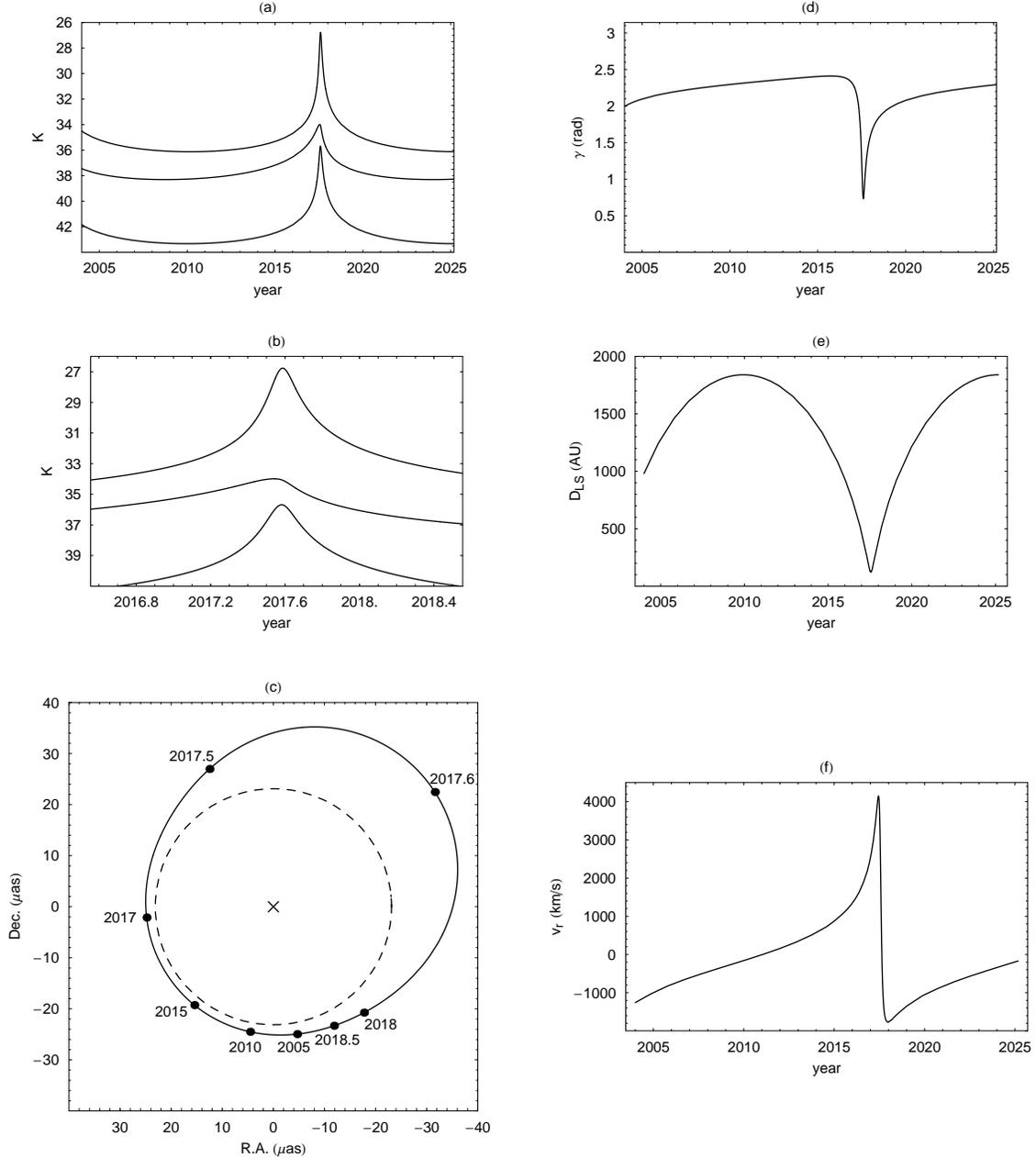}}
 \caption{$S2.$ -- (a)
Light curves for the gravitational lensing images of S2. From top
to bottom, we have the secondary, the third order and fourth order
images. (b) Details of the same light curves at the peak epoch.
(c) Trajectory of the secondary image around the central black
hole (marked by the cross in the middle). The dashed circle is its
apparent horizon. (d) The angle $\gamma$, formed by the line
connecting S2 with the central black hole and the optical axis;
(e) distance between S2 and Sgr A*; (f) radial velocity of S2. All
quantities are plotted as functions of time. }
 \label{Fig S2}
\end{figure*}

$S2.$ -- The spectral analysis of S2 reveals it as a main sequence
star of 15 M$_\odot$ of class O8 -- B0 \citep{Ghez2,Eis}. It is
also famous as the star providing the best orbital constraint on
the mass of the central black hole.

Figure \ref{Fig S2}a shows the apparent magnitudes in the K-band
of its most important gravitational lensing images as functions of
time, throughout the next orbital period. From top to bottom, we
have the secondary image, the third order and the fourth order
one. Their brightness changes very slowly during the whole orbital
period, save for a sharp peak corresponding to the periapse epoch,
as it is evident from Figure \ref{Fig S2}e, which shows $D_{LS}$
as a function of time. This is easily understood as a consequence
of the $D_{LS}^{-2}$ dependence in the magnification formula
(\ref{mu}). In general, because of this dependence, we can say
that highly eccentric orbits have more prominent peaks (once the
inclination of the orbit has been fixed), since the minimum
distance of the star from the black hole becomes smaller and
smaller with respect to the maximum distance. Also the duration of
the peak is determined by the eccentricity, in that the time the
star spends to complete the passage at the periapse is shorter for
more eccentric orbits.

To be more quantitative, it is possible to show that for high
eccentricities ($1-e \ll 1$), the ratio between the duration of
the peak and the orbital period scales as $1-e$, while the ratio
between the maximum and minimum brightness scales as $(1-e)^{-2}$.

Figure \ref{Fig S2}b shows a detail of the peak epoch. We see that
the new estimates for the orbital parameters of S2 bring the
secondary image below K=27 in the middle of 2017. We recall that
the previous estimates for the peak magnitude gave K=30
\citep{DeP,BozMan} for the beginning of 2018. The third order
image reaches K=34 at the same time, while the fourth order image
has K=36.

In Figure \ref{Fig S2}c we report the trajectory of the secondary
image around the central black hole, marked by a cross. The
apparent event horizon is the circle of angular radius $\theta_m$
shown in dashed style. The trajectory of the secondary image is an
ovoidal figure very close to the apparent event horizon for a
large fraction of the orbital period. The distance of the image
from the position of the central black hole is fixed by the lens
equation (\ref{Lens Eq}). It basically depends on the angle
$\gamma$, formed by the line joining the source to the lens with
the optical axis. To see this correspondence in the specific case
of S2, one can compare with Figure \ref{Fig S2}d, where $\gamma$
is plotted as a function of time. We see that during the periapse
epoch, $\gamma$ becomes smaller than $\pi/2$, which means that S2
is behind the black hole, the minimum value being 0.83 rad. At
this epoch, the secondary image reaches the largest distance from
the black hole. By contrast, during most of the period S2 is on
the same side of the observer with respect to the black hole.
During this period, the secondary image comes closer to the event
horizon, since the photons require a larger deflection by the
black hole in order to reach the observer.

The position angle of the secondary image around the black hole is
always opposite to the position of the direct image, in the
Schwarzschild black hole hypothesis. The third and fourth order
images are formed very close to the apparent event horizon (their
typical separation is a fraction of $\mu$as), so that in the scale
of Figure \ref{Fig S2}c they would just be superposed on the
dashed circle. Therefore, we omit them in all figures. For
completeness, we have also included the plot of the radial
velocity (Fig. \ref{Fig S2}f).

\begin{figure*}
\resizebox{\hsize}{!}{\includegraphics{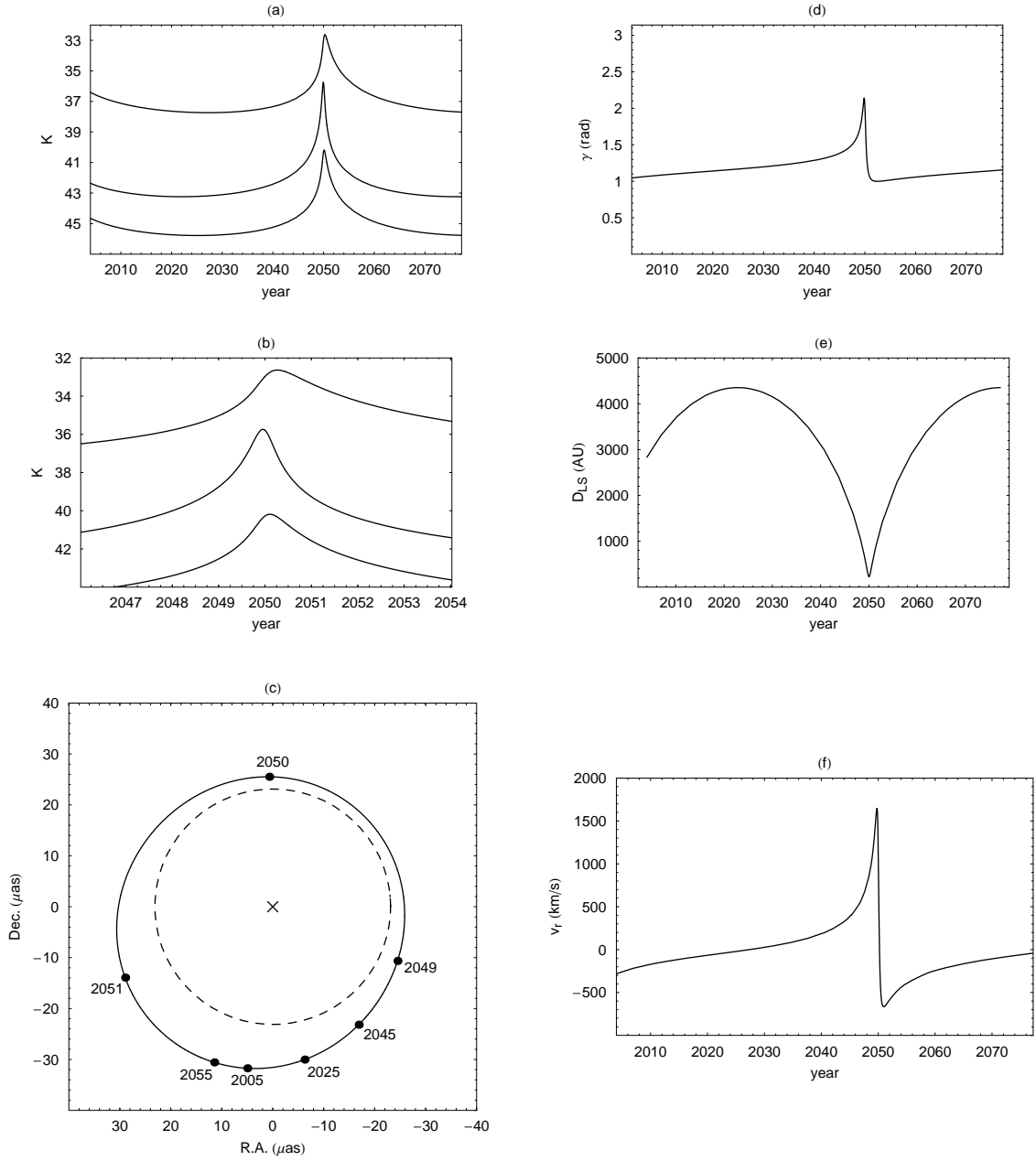}}
 \caption{Results for S12. For the meaning of each panel, refer to the caption of Fig. \ref{Fig S2}.
 }
 \label{Fig S12}
\end{figure*}

$S12.$ -- The results for this star are summarized in Figure
\ref{Fig S12}. Here the maximal brightness is reached in
{\scriptsize C.E.} 2050, when S12 will reach its minimum distance
from Sgr A* again. At that time S12 will be in front of the black
hole ($\gamma>\pi/2$), as can be seen from Figure \ref{Fig S12}d.
Comparing the light curves for the images of S12 with those of S2,
we find a very interesting difference. The magnification peaks are
determined not only by the minimum $D_{LS}$, but also by the
degree of alignment of the star with the optical axis. The peaks
of the negative parity images (secondary and fourth order) are
enhanced when $\gamma$ is closer to 0, i.e., when the star lies
behind the black hole. The third order image and the other
positive parity images are maximally magnified when $\gamma$ is
closer to $\pi$, i.e., when the star lies in front of the black
hole. Since S2 is behind the black hole at the periapse, the
secondary and fourth order images are strongly enhanced, while the
third order image has only a quite small peak. The brightness of
the fourth order image almost reaches that of the third order
image. On the contrary, S12 is in front of the black hole at the
periapse, and the third order image is more enhanced than the
secondary and the fourth order image. Its brightness almost
reaches that of the secondary image. Moreover, since the minimum
value of $\gamma$ is reached just few months after the periapse,
the peaks of the secondary and of the fourth order images are
broadened and "attracted" toward the time at which $\gamma$ is
minimal. The same happens to the third order image of S2, which is
stretched toward earlier times, when $\gamma$ is closer to $\pi$.
A radical difference with respect to S2 also arises in the
trajectory of the secondary image. Looking at Figure \ref{Fig
S12}c, we see that the secondary image spends most of the time at
larger distances from the black hole, since S12 is behind the
black hole for most of its orbit (Fig. \ref{Fig S12}d). During the
periapse, S12 very quickly passes in front of the black hole and
the secondary image grazes the apparent horizon making half of the
tour in a very short time. This situation is the opposite of that
of S2, which is in front of the black hole for most its orbit and
behind it just at the periapse. Another interesting thing to
notice is that the secondary image of S12 does not move very far
from the apparent horizon, since the orbit has a lower inclination
and $\gamma$ is bounded to a lower range around $\pi/2$.

\begin{figure*}
\resizebox{\hsize}{!}{\includegraphics{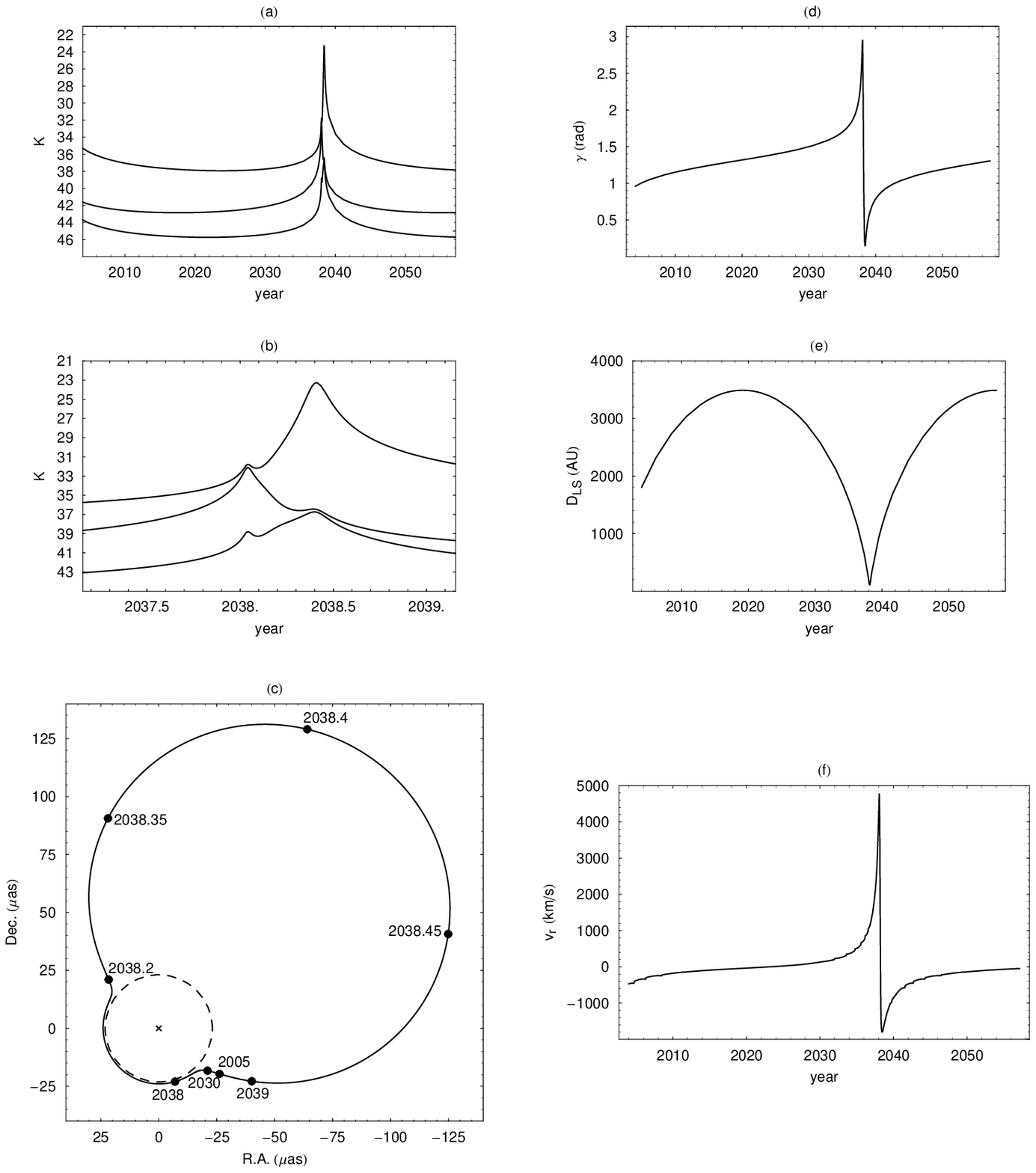}}
 \caption{Results for S14. For the meaning of each panel, refer to the caption of Fig. \ref{Fig S2}.
 }
 \label{Fig S14}
\end{figure*}

$S14.$ -- This star represents the most interesting lensing
candidate, since its orbit is almost edge-on ($i=97.3$ deg), and
its eccentricity brings it very close to the black hole. The best
alignment and anti-alignment times are very close to the periapse
epoch and are responsible for two subpeaks, which are unresolved
in Figure \ref{Fig S14}a, but clearly visible in Figure \ref{Fig
S14}b, which zooms on the peak. The periapse takes place while
$\gamma \sim \pi/2$, so that this represents an intermediate case
with respect to the former ones. First we have the anti-alignment
peak, mostly enhancing the third order image, which reaches the
brightness of the secondary image. Then we have the alignment
peak, which mostly enhances the secondary and fourth order images.
If the periapse coincided with the best alignment or
anti-alignment time, the corresponding peak would have been much
more prominent than shown in Figure \ref{Fig S14}b.

What makes S14 particularly interesting is the high brightness
attained by the secondary image. At the best alignment time the
secondary image has K=23. This is because we are close to a weak
field gravitational lensing situation as can also be seen by the
analysis of Figure \ref{Fig S14}c. The trajectory of the secondary
image is composed by two half rings: one grazing the apparent
horizon when S14 is in front of the black hole and the other being
the approximate weak field solution for the secondary image in a
typical microlensing event. In fact, consider a source passing
behind a point-lens along the line parameterized by
\begin{equation}
\vec{y}=(v t, b_0),
\end{equation}
where $v$ is the transverse velocity and $b_0$ is the microlensing
impact parameter. Then the approximate solution for the secondary
image in the case $b_0 \gg 1$ is
\begin{equation}
\vec{x}\simeq \frac{1}{b_0^2+v^2t^2}(-v t, -b_0),
\end{equation}
which parameterizes a circle of radius $1/2b_0$ centered at
$(0,-1/2b_0)$. This is exactly the big circle described by the
secondary image in Figure \ref{Fig S14}c, apart from small
distortions caused by the curvature of the orbit of S14.

The duration of the main peak is about one month and is determined
by the velocity of S14 at the periapse epoch. One may wonder
whether the direct image is affected in a significant way by this
almost weak field event. However, we find $\Delta \mu/\mu\sim
10^{-3}$ for the direct image, which makes the peak practically
unobservable. The next brightness peak of the secondary image of
S14 will occur in {\scriptsize C.E.} 2038, according to the
current estimates of the orbital parameters, and the maximal
angular distance from the apparent event horizon is about 0.125
mas. The perspectives for the observation of such event will be
discussed in the next section.

The third order image reaches K=32, deserving particular
attention, since the eventual observation of such image would be
of striking importance for the physics of the gravitational field
in the strong regime.

\begin{figure*}
\resizebox{\hsize}{!}{\includegraphics{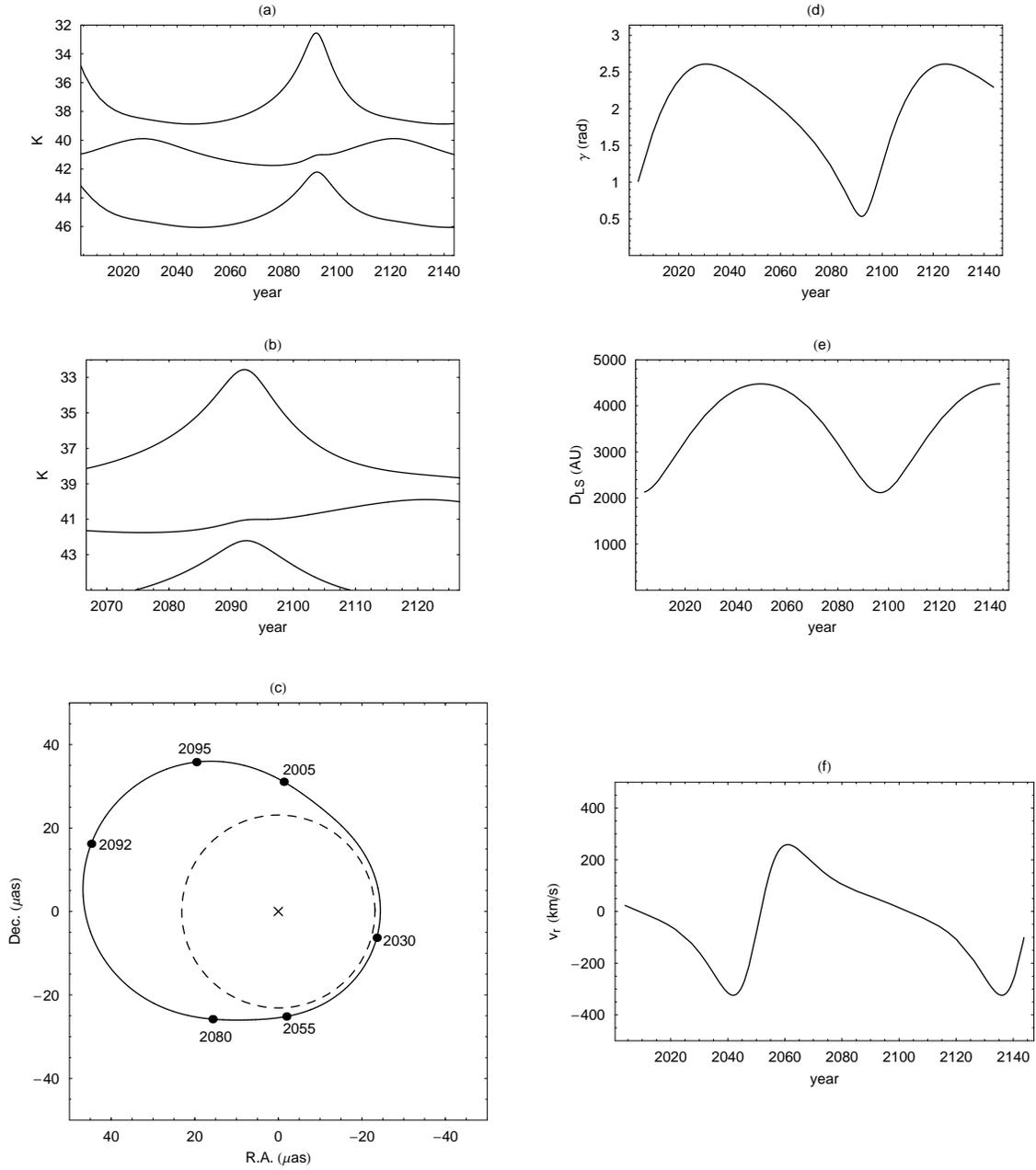}}
 \caption{Results for the star S1. For the meaning of each panel, refer to the caption of Fig. \ref{Fig S2}.
 }
 \label{Fig S1}
\end{figure*}

$S1.$ -- This star is characterized by a lower eccentricity, as
clearly visualized in Figure \ref{Fig S1}e, where the range of
values of $D_{LS}$ is small, and in Figure \ref{Fig S1}f, where we
see low radial velocities compared to the other stars. It happens
that the periapse is in coincidence with the best alignment time,
(see Fig. \ref{Fig S1}d), and in fact the peaks in the secondary
and fourth order images are very enhanced, while the third order
image has only a small modulation throughout the orbit (Fig.
\ref{Fig S1}a). We also observe that the third order image has a
small secondary peak at the best alignment time, which is formed
in a way analogous to that discussed for S14. In any event, the
secondary image will stay fainter than K=32 at the next peak,
occurring only in {\scriptsize C.E.} 2091, so that this is a less
interesting case for observations.

The trajectory of the secondary image (Fig. \ref{Fig S1}c) can be
easily interpreted from the considerations given in the previous
cases.

\begin{figure*}
\resizebox{\hsize}{!}{\includegraphics{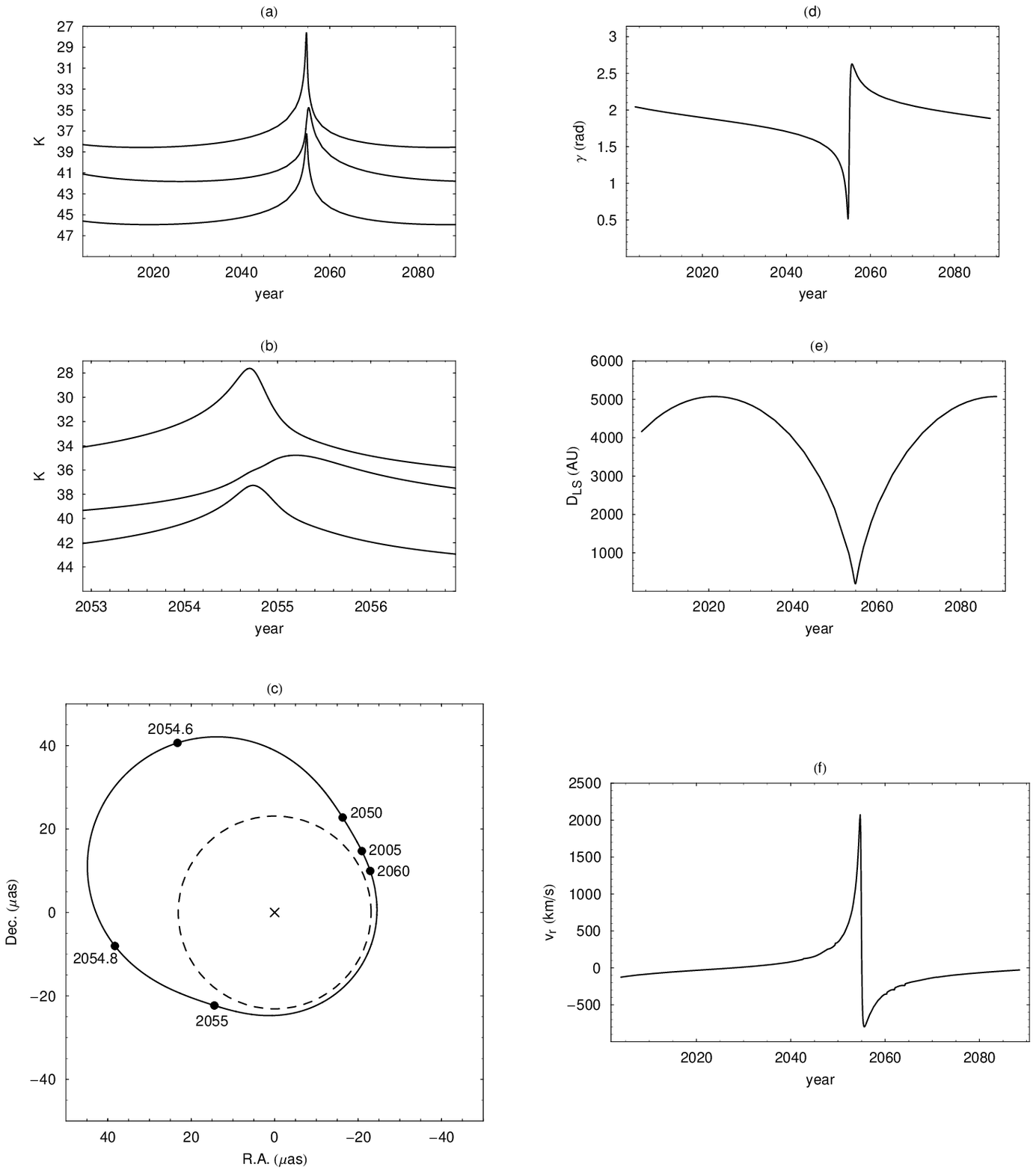}}
 \caption{Results for S8. For the meaning of each panel, refer to the caption of Fig. \ref{Fig S2}.
 }
 \label{Fig S8}
\end{figure*}

$S8.$ -- This star is known to have a high eccentricity (so high
that, before the last data were released, it was still unclear
whether the orbit was open or closed \citep{Ghez3}). According to
\citet{Eis}, the orbit is closed and it makes sense to discuss
predictions for the next brightness peaks. Even though the
periapse distance is small, the inclination is not very high, so
that we do not have the same favorable situation as S14. In
{\scriptsize C.E.} 2054, the secondary image will "only" reach
K=28. The periapse lies between the best alignment and
anti-alignment times, being a bit closer to the first
configuration. Then the secondary and fourth order image peaks are
a bit more prominent than the peak of the third order image, which
also shows a very small deformation at the best alignment time.

\begin{figure*}
\resizebox{\hsize}{!}{\includegraphics{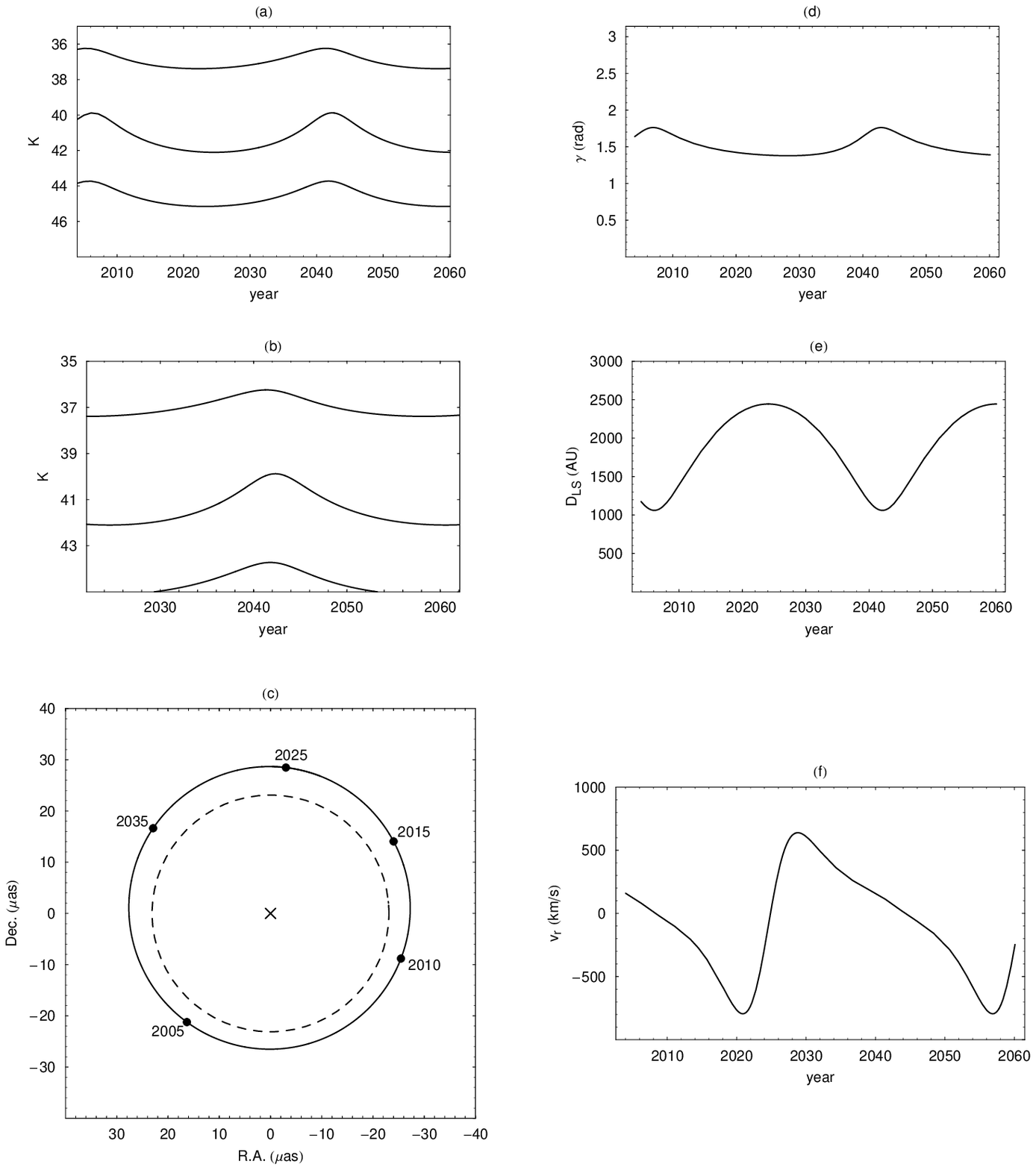}}
 \caption{Results for S13. For the meaning of each panel, refer to the caption of Fig. \ref{Fig S2}.
 }
 \label{Fig S13}
\end{figure*}

$S13.$ -- For this star the uncertainty on the orientation of the
orbital plane remains very large. So, our predictions can just
give an indication, which will be updated when further
measurements are available. In any case, the inclination of the
orbital plane should be quite low and the images have a nearly
constant low brightness, as shown in Figure \ref{Fig S13}a. The
distance of the secondary image from the black hole is also nearly
constant. Hence, this star also seem less promising as a candidate
for gravitational lensing observations.

Table 2 summarizes the main characteristics of the gravitational
lensing images for our six stars, allowing an immediate reading of
all results.

\begin{table*}
\centering

\begin{tabular}{ccccccc}
 \hline \hline

    & S2 & S12 & S14 & S1 & S8 & S13 \\
  \hline
  $K_2$ (mag) \dotfill & 26.8 & 32.6& 23.3 & 32.6 & 27.6 & 36.2\\
  $T_2$ (yr)  \dotfill & 2017.59& 2050.27 & 2038.41 & 2092.13 & 2054.7 & 2041.38\\
  $\Delta \theta_2$ ($\mu$as) \dotfill & 16 & 9.8& 125 & 25 & 26 & 5.6 \\
  $K_3$ (mag) \dotfill & 34.0& 35.7 & 32.1 & 39.9& 34.8& 39.9\\
  $T_3$ (yr)  \dotfill & 2017.54& 2049.95 & 2038.04 & 2121.29& 2055.2 & 2042.33\\
  $\Delta \theta_3$ ($\mu$as) \dotfill & 0.33 & 0.25 & 0.62 & 0.41 & 0.42 & 0.17\\
  $K_4$ (mag) \dotfill & 35.7& 40.2 & 36.7 & 42.2& 37.3& 43.7\\
  $T_4$ (yr)  \dotfill & 2017.58& 2050.11 & 2038.4 & 2092.44& 2054.74 & 2041.8\\
  $\Delta \theta_4$ ($\mu$as) \dotfill & 0.014 & 0.011 & 0.026 & 0.017 & 0.017 &  0.007\\

 \hline

\end{tabular}
\caption{Summary of the main features of the relativistic images
of the six stars examined in this paper. $K_i$ is the peak K-band
magnitude for the $i$-th image; $T_i$ is the time of the peak;
$\Delta \theta_i=\theta_i - \theta_m$ is the maximal angular
distance from the apparent event horizon.}
\end{table*}

\section{Perspectives for observations}

From our analysis, S14 emerges as the most interesting lensing
candidate, because of its small minimum distance from the black
hole and because of the high inclination of its orbit, which
allows good alignments and anti-alignments with Sgr A*. The
secondary image will reach K=23 at the next periapse epoch, which
will occur around {\scriptsize C.E.} 2038. The rapid motion of
this image would help to distinguish it from background sources of
similar brightness, while its distortion should also be an
unambiguous sign of its gravitational lensing nature with respect
to fast moving stars around Sgr A*. The theoretical baseline
needed to resolve two sources of comparable brightness placed at
an angular distance of $\theta=0.12$ mas in the K-band, amounts to
$\lambda/ \theta=3.8$ km. However, the apparent magnitude of Sgr
A* is K=18.8 \citep{Ghez4,Genzel2}, which means that we have to
distinguish a faint source from a 63 times brighter contaminant.
This complication can be tackled by longer baselines and nulling
interferometry techniques. Sgr A* is still unresolved in this
band. One may be worried that if its infrared flux comes from a
region of several Schwarzschild radii, then the secondary images
of lensed stars would be merged within its diffuse infrared
emission. However, current models explaining the rapid observed
variability suggest that this emission takes place within 10
$R_{Sch}$ from the central black hole \citep{YQN}. In this case,
the secondary image of S14 would come out of the infrared emission
of Sgr A* and could be resolved, in principle, by interferometry
facilities.

At the present time, the best resolution in the K-band is achieved
by the Keck telescope and the VLT units, the latter reaching 75
mas without interferometry, thanks to active and adaptive optics.
The VLT units can be combined to perform interferometry
observations with an equivalent baseline of 200 m and a maximal
angular resolution of 2.2 mas (http://www.eso.org/projects/vlti).
A new interferometric instrument under construction is the Large
Binocular Telescope (LBT), which will use two 8.4 m telescopes
reaching an equivalent resolution of a 22 m telescope
(http://medusa.as.arizona.edu/lbto).

New perspectives are opened by space telescopes. The Terrestrial
Planet Finder (TPF) will be composed by four 3.5 m telescopes
working in team. The design details are still under study; in
particular, the four units may be mounted on the same structure or
flying in formation several meters apart
(http://www.terrestrial-planet-finder.com). A similar project is
DARWIN, which will exploit six telescopes flying in formation with
an equivalent aperture of 50 meters
(http://ast.star.rl.ac.uk/darwin). These two missions are foreseen
to launch around 2015 and are mainly aimed at exoplanet
researches. The problem of resolving a faint planet from its
mother-star, which is typically millions of times brighter, can be
faced by the technique of nulling interferometry, which combines
the signal from a number of different telescopes in such a way
that the light from the central star is cancelled out, leaving the
much fainter planet easier to see. Of course the situation of a
faint gravitational lensing image close to Sgr A* has great
analogies with that of exoplanets and the same techniques can be
used. In order to achieve resolutions better than 0.1 mas, we need
a baseline larger than several kilometers, which must be kept
stable at a very high precision. A similar problem is faced by the
LISA mission in the search for gravitational waves. In this
mission, three spacecraft should fly in a triangular formation at
relative distances of 5 million km (http://lisa.nasa.gov).
Finally, the MAXIM mission will perform X-ray interferometry with
a constellation of several spacecraft and a detector kept 450 km
behind the mirrors (http://maxim.gsfc.nasa.gov). To accomplish
formation flying with high accuracy, these missions are planning
to use a laser ranging system between spacecraft and
microthrusters to offset drifts.

Given the present situation and all the technical advances that
will soon be exploited in scientific researches, 0.1 mas
resolution in the K-band and the detection of secondary
gravitational lensing images around Sgr A* seem just the next step
after the generation of telescopes currently under study.
Hopefully, new instruments will be available before the next
periapse passage of S14 is in {\scriptsize C.E.} 2038, to catch
its secondary image. For S2 and S8 the situation is more
difficult, since the peak magnitude and the angular separations of
the secondary images are lower, but the numbers do not rule out
the possibility of observing them with reasonable future
facilities.

As regards higher order images, S14 is still the best candidate.
However, the peak magnitude of the third order image is K=32.1,
and the separation from the apparent horizon is only 0.62 $\mu$as.
In order to see such a faint image so close to the event horizon
we would need a complete imaging at very high resolution ($<0.1$
$\mu$as) of Sgr A*. In the K-band this seems a very difficult
task. The highest resolution ever reached is 18 $\mu$as in the mm
band of the electromagnetic spectrum \citep{Krich}. Higher
resolution imaging seems to be possible at sub-mm wavelengths
\citep{FMA}, though no compact radio sources (suitable for
gravitational lensing) are known around Sgr A*. Finally, MAXIM
claims that a 0.1 $\mu$as resolution is achievable in the X-ray
band. In this case, one should look for suitable X-ray sources
around the Galactic center, which could be provided by a recently
observed population of black holes within 1 pc from Sgr A*
\citep{Muno}. Higher order images thus stand as a possible
long-term target for future observations.

\section{Discussion and Conclusions}

It is really amazing that at present time it makes sense to
discuss on gravitational lensing in strong fields by real physical
objects. Indeed, the existence of a supermassive black hole at the
center of our Galaxy with its rich stellar environment provides a
unique laboratory for this research. However, this would not be
enough without the great technical advances in high resolution
observations of the last years, which allow us to speak about
microarcsecond resolutions in sub-mm observations and possibly in
higher frequency bands. The proper motion observations of stars
surrounding Sgr A* are progressing daily, providing more and more
precise estimates for the orbital parameters. In this work, we
have analyzed six stars enjoying a good determination of their
orbital parameters. Our aim has been to look for observable
gravitational lensing effects on these stars by the black hole in
Sgr A*. Our investigation has considered the full set of
gravitational lensing images, including weak field and
relativistic ones, the latter being of striking importance for a
deeper understanding of General Relativity, as they provide
observational tests for gravitational theories in previously
unprobed regimes. We have explicitly shown the light curves of the
secondary, third order and fourth order images for our six stars.
Moreover, we have shown the position in time of the secondary
image w.r.t. the central black hole. We have given physical
interpretations to our results, relating, in particular, the peaks
to minima of $D_{LS}$ and best alignment and anti-alignment times.

We want to stress that this work is just a first analysis, which
is susceptible to be corrected by eventual upcoming estimates of
the orbital parameters, enlarged to eventual new candidate stars,
and generalized to different classes of black holes. The most
physically motivated generalization of the Schwarzschild black
hole is of course the Kerr solution. However, while in the
Schwarzschild case the deflection angle can be still expressed
analytically by equation (\ref{alpha exact}), the treatment of
gravitational lensing by Kerr black holes is done in a completely
numerical way at the present time \citep{CunBar,Vie,RauBla}.
Moreover, the exact value of the spin of Sgr A* is still very
uncertain and inferred by indirect deductions \citep{LiuMel,Asc}.
The deviations from the Schwarzschild predictions are bigger for
higher order images, while they become very small in the weak
field limit. This means that a high value of the spin would
considerably alter the predictions for higher order images, but,
for example, the peak of the secondary image of S14 would be
affected in a negligible way. So, for a first analysis,
Schwarzschild lensing is basically sufficient to understand the
highlights of the lensing scenario and the order of magnitude of
the observables. Kerr lensing comes as a second step refinement.

Another topic that deserves some discussion concerns the
uncertainties in our predictions coming from the present
uncertainties in the estimates of the orbital parameters. The
first thing to notice is that all these stars have been observed
during their periapse epochs. We can thus be confident that even
if the semimajor axis and the eccentricity are affected by large
errors, the periapse distance is well determined by direct
observations. Since the inclinations are also relatively well
determined, we can conclude that the magnitudes of the peaks
cannot be very different from the ones that we have calculated
using the best-fit values of the orbital parameters. By contrast,
the precise timing of the peaks is strictly related to the
accuracy of the orbital period estimate. This can be established
only with a good sampling on the whole orbit which requires more
systematic observations in the future. As an example, given the
present uncertainties, the brightness peak of the images of S14
could be displaced few years later or before the date that we have
indicated using the best-fit value.

Finally, with the progress of the observation techniques, it is
inevitable that new stars will be discovered around Sgr A* and
followed during their orbits. If new stars are discovered closer
and closer to Sgr A*, then the thin lens approximation cannot be
used any longer. In that case, the analysis should proceed in the
way indicated by \citet{CunBar}. All these considerations make us
confident that even better candidates than S14 may be found in the
future to open the era of gravitational lensing by strong fields.

\begin{acknowledgements}
We thank Gaetano Scarpetta for useful comments on the manuscript.
We also thank the referee for many suggestions which have helped
us to improve our paper considerably.
\end{acknowledgements}

------------------------------------------------------------------

\bibliographystyle{aa}

\end{document}